\documentclass[fleqn,usenatbib]{mnras}

\usepackage[T1]{fontenc}
\usepackage{xcolor}
\usepackage[normalem]{ulem}

\DeclareRobustCommand{\VAN}[3]{#2}
\let\VANthebibliography\thebibliography
\def\thebibliography{\DeclareRobustCommand{\VAN}[3]{##3}\VANthebibliography}

\usepackage{graphicx}	
\usepackage{amsmath}	
\usepackage{amssymb}	

\title[Active Reflector Technology in FAST-like telescopes]{MNRAS \LaTeXe\\
Adapting Active Reflector Technology for greater sensitivity and sky-coverage in FAST-like Telescopes}

\author[Jian-Ling Li,Bo Peng,Cheng-Jin Jin et al.]{
Jian-Ling Li,$^{1,2,3}$\thanks{E-mail:\href{mailto:jlli@nao.cas.cn}{jlli@nao.cas.cn}}
Bo Peng,$^{1,2}$\thanks{E-mail:\href{mailto:pb@nao.cas.cn}{pb@nao.cas.cn}}
Cheng-Jin Jin$^{1,2}$
Hui Li$^{1,2}$
Richard G. Strom$^{4,5}$
Bin Liu$^{1,2}$\thanks{E-mail:\href{mailto:bliu@nao.cas.cn}{bliu@nao.cas.cn}}
Xiao-Ming Chai$^{1,2}$
\and Li-Jia Liu$^{1,2,3}$
\\
$^{1}$National Astronomical Observatories, Chinese Academy of Sciences,20A Datun Road, Chaoyang District, Beijing, 100101, China\\
$^{2}$CAS Key Laboratory of FAST, National Astronomical Observatories, Chinese Academy of Sciences,\\
    20A Datun Road, Chaoyang District, Beijing, 100101, China\\
$^{3}$University of Chinese Academy of Sciences, No.19(A) Yuquan Road, Shijingshan District, Beijing,100049, China\\
$^{4}$Anton Pannekoek Institute for Astronomy, Science Park 904, University of Amsterdam, 1098 XH Amsterdam, The Netherlands\\
$^{5}$ASTRON, Radio Observatory, Oude Hoogeveensedijk 4, 7991 PD Dwingeloo, The Netherlands
}

\date{Accepted XXX. Received YYY; in original form ZZZ}

\pubyear{2015}

\begin{document}
\label{firstpage}
\pagerange{\pageref{firstpage}--\pageref{lastpage}}
\maketitle

\begin{abstract}
The Five-hundred-meter Aperture Spherical radio Telescope (FAST), the largest single dish radio telescope in the world, has implemented an innovative technology for its huge reflector, which changes the shape of the primary reflector from spherical to that of a paraboloid of 300\,m aperture. Here we explore how the current FAST sensitivity can potentially be further improved by increasing the illuminated area (i.e., the aperture of the paraboloid embedded in the spherical surface). Alternatively, the maximum zenith angle can be increased to give greater sky coverage by decreasing the illuminated aperture.Different parabolic apertures within the FAST capability are analyzed in terms of how far the spherical surface would have to move to approximate a paraboloid. The sensitivity of FAST can be improved by approximately 10\% if the aperture of the paraboloid is increased from 300\,m to 315\,m. The parabolic aperture lies within the main spherical surface and does not extend beyond its edge. The maximum zenith angle can be increased to approximately 35 degrees from 26.4 degrees, if we decrease the aperture of the paraboloid to 220\,m. This would still give a sensitivity similar to the Arecibo 305\,m radio telescope. Radial deviations between paraboloids of different apertures and the spherical surfaces of differing radii are also investigated. Maximum zenith angles corresponding to different apertures of the paraboloid are further derived. A spherical surface with a different radius can provide a reference baseline for shape-changing applied through active reflector technology to FAST-like telescopes.
\end{abstract}

\begin{keywords}
telescopes -- methods: analytical -- instrumentation: adaptive optics--methods: observational
\end{keywords}



\section{Introduction}

The world\arcmin s largest single dish antenna, the Five-hundred-meter Aperture Spherical radio Telescope (FAST), has three outstanding features \citep{2006ScChG..49..129N}: a unique karst depression found in southern Guizhou province as its site \citep{1996ltwg.conf...59N}, an active primary reflector which directly corrects for spherical aberration \citep{1998MNRAS.301..827Q}, and a light-weight feed cabin driven by cables and a servomechanism\citep*{1996ltwg.conf..152D} plus a parallel robot as a secondary adjustable system to precisely position the feeds. The main reflector of FAST is a spherical bowl with a radius of 300 m and an opening up to 500 m in diameter \citep{2000ASPC..213..523N,2003AcASn..44S..13N}. The illuminated part of the main spherical reflector is to be continuously adjustable to conform to a paraboloid (\citealt{2009IEEEP..97.1391P}; \citealt*{2000SPIE.4015...45P}; \citealt{2002RaScB.300...12P}). To achieve this, active reflector technology has been adopted for FAST. It is the most significant difference between FAST and Arecibo, and is the key innovation in FAST \citep*{2020RAA....20....1S}. The construction of FAST was completed in September, 2016, after which FAST entered a commissioning phase, and officially opened for operation in January, 2020.

\begin{figure}
	\includegraphics[width=\columnwidth]{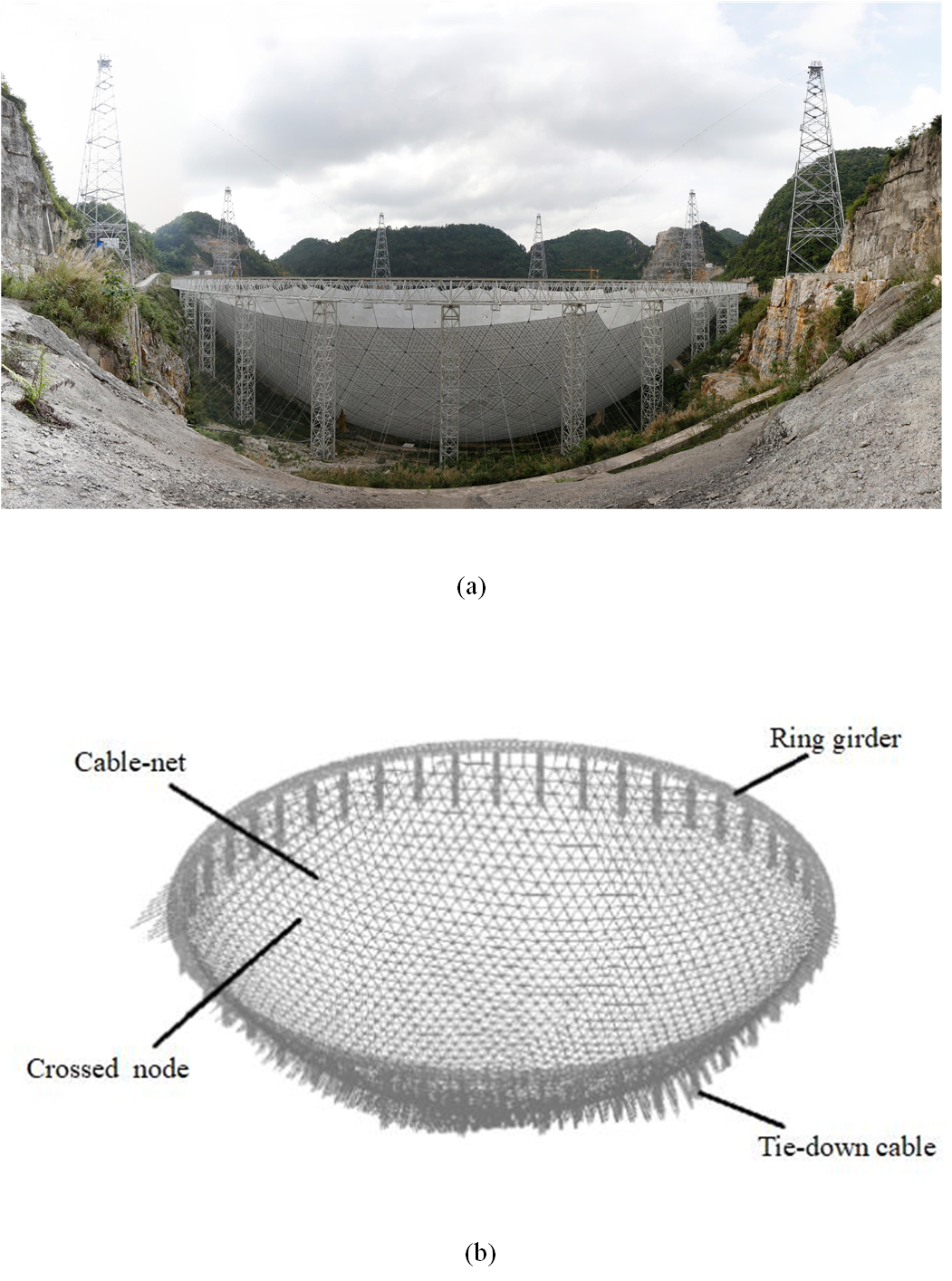}
    \caption{The FAST photo taken with a wide-angle lens (a) and its dish structure(b)}
    \label{fig:fig1 figure}
\end{figure}

The active reflector system of FAST is mainly composed of a ring girder, a cable-net, tie-down cables, actuators, reflector elements and ground anchors \citep{2017Key}. The FAST cable-net mainly comprises 6670 steel cables and ~2225 crossed nodes \citep{2019SCPMA..6259502J}. The crossed nodes are used as control points, which are tied to the actuators by tie-down cables \citep{doi:10.1142/S0218271811019335}. The main structure of the active reflector is shown in Figure ~\ref{fig:fig1 figure}. The shape of the illuminated area can be changed into a paraboloid of 300\,m aperture by control of the actuators.

The FAST receivers were designed to illuminate a 300\,m parabolic aperture, covering the frequency range between 70\,MHz and 3\,GHz, among which a multi-horn receiver of 19 beams at L band is currently mounted on the focus cabin, whose edge taper is between  -7 dB at lower frequency end and -12 dB  at higher frequency end.

\begin{figure}
	\includegraphics[scale=0.8, angle=0]{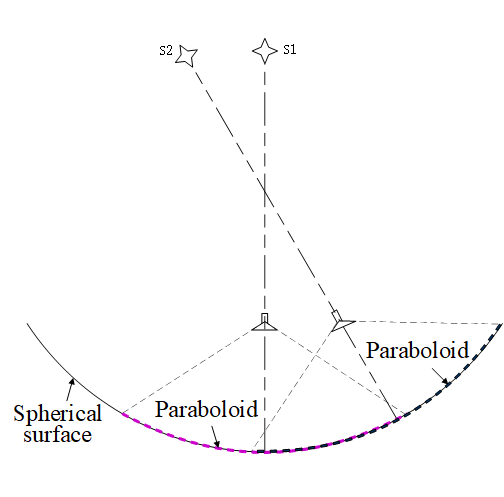}
    \caption{Schematic diagram of FAST observation principle}
    \label{fig:fig2_figure}
\end{figure}

During an observation, the illuminated area moves within the main spherical reflector (except in the drift scan mode, for example). A schematic diagram of the FAST observation principle is shown in Figure ~\ref{fig:fig2_figure}. A study of shape change in going from the spherical surface to a paraboloid is one of the key points for the active reflector of the telescope. The effective collecting area correspondingly increases when the aperture of the paraboloid increases, which means the sensitivity of the telescope is improved. The maximum zenith angle increases when the aperture of the paraboloid decreases, which means the sky coverage of the telescope becomes larger. A larger sky coverage is important for FAST ($l = +25.6^{\circ}$) to observe the Orion region ($\delta \sim -5^{\circ}$), or Cas A ($\delta \sim +60^{\circ}$),the brightest source at centimeter wavelengths.

For example, for FAST, if the aperture of the paraboloid is increased to 315\,m, the sensitivity of the telescope will be improved by approximately 10\%, (assuming the noise temperature, aperture efficiency, observing time, etc., are unchanged.). The paraboloid stays within the main spherical surface and does not extend beyond the edge of the spherical primary. If the aperture of the paraboloid is decreased to 205\,m, the maximum zenith angle achievable will increase by approximately 10 degrees from current limit 26.4 degree (corresponding to 40\% of the celestial sphere) to 36.4 degree (53\% sky).

Application of the active reflector technology to spherical primary reflectors of different radii is further analyzed. Such analysis can be applied to the proposed expanded array of FAST-like telescopes. By contrast with the Square Kilometre Array (SKA), a technical proposal to exploit the karst depressions in Guizhou to build a LDSN (Large Diameter Small Number) radio telescope array was proposed in 1994  \citep{1997hsra.book..278P,1998IAUS..179...93P}. An expanded array of FAST-like telescopes will greatly enhance both sensitivity and resolution.

In the next section, the basic calculation and recommended calculation conditions for the paraboloid are given. The analysis of the FAST paraboloids of differing apertures is provided in section 3, and this study suggests the potential improvement in its performance. Further analyses of the main spherical surface for different radii and for paraboloids of different apertures are given in section 4. This analysis is applicable to FAST-like radio telescopes generally. In section 5, the theoretical calculation results are discussed and a summary is given.

\section{Analysis of the paraboloid and the spherical surface}

\begin{figure}
	\includegraphics[scale=0.6, angle=0]{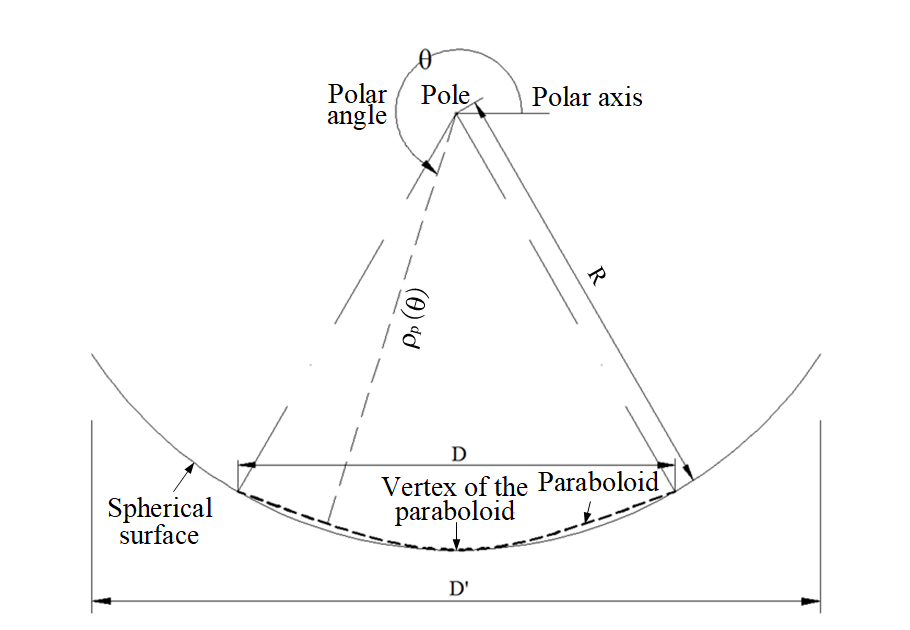}
    \caption{Diagram of the paraboloid and the spherical surface in polar coordinates. The solid line represents the spherical surface, the dashed line represents the paraboloid shaped by distorting the spherical surface; the vertex of the paraboloid is also shown. The aperture of the paraboloid is denoted by $D(m)$, the radius of the spherical surface is $R(m)$ , and the opening aperture of the spherical surface is given by $D'(m)$.$\rho_p$ is the radial length from the pole to the paraboloid in meters,which is a function of polar angle($\theta$).}
    \label{fig:fig3_figure}
\end{figure}

The spatial relationship between the paraboloid and the spherical surface is analyzed. The paraboloid is analyzed and calculated in the radial direction of the spherical surface. The center of the spherical surface is taken as the pole (origin), and a horizontal line from the pole is taken as the polar axis, whereby polar coordinates can be established. The paraboloid at the bottom of the spherical surface is considered to be a parabola in the specified position in these polar coordinates. Our interest is in the radial deviation of the paraboloid from the spherical surface. A diagram of the paraboloid and the spherical surface is shown in Figure ~\ref{fig:fig3_figure}.

We set $\rho_s$  as the radial length from the pole to the spherical surface in meters, $\rho_p$ is the radial length from the pole to the paraboloid in meters.
The formula representing the spherical surface in this polar coordinate can be expressed as formula ~(\ref{eq:eq1}),

\begin{equation}
    \rho_s=R.
	\label{eq:eq1}
\end{equation}

Due to the rotational symmetry of the paraboloid and the spherical surface, the relationship between the paraboloid in different positions and the main spherical surface is the same. The formula representing the paraboloid in polar coordinates can be expressed as follows,

\begin{equation}
    \rho_p^2\cos^2\theta=2p\rho_p\sin\theta+c,
	\label{eq:eq2}
\end{equation}
where $\theta$ stands for the polar angle of the paraboloid in radians, $\frac{p}{2}$ is the focal length of the paraboloid in meters, and $c$ is a constant related to $p$.

When the paraboloid is in the specified position as shown in Figure~\ref{fig:fig3_figure}, the polar angle range of the paraboloid corresponding to its aperture can be expressed as $(\frac{3}{2}\pi-\arcsin(\frac{0.5D}{R}), \frac{3}{2}\pi+\arcsin(\frac{0.5D}{R}))$. The outer edge of the illuminated aperture coincides with the main spherical surface \citep*{article}. If this is used as the basic constraint when analyzing the paraboloid, then in the specified position it coincides with $(R,\frac{3}{2}\pi+\arcsin(\frac{0.5D}{R}))$, and formula ~(\ref{eq:eq3}) can be obtained,

\begin{equation}
    c=R^2\cos^2(\frac{3}{2}\pi+\arcsin(\frac{0.5D}{R}))-2pR\sin(\frac{3}{2}\pi+\arcsin(\frac{0.5D}{R})).
	\label{eq:eq3}
\end{equation}

Using the basic constraint mentioned above, the following two conditions are the main ones adopted as calculation and selection prerequisites for the paraboloid:

(1)	Select the paraboloid with the minimum $d$ at its vertex, where $d$ is the radial deviation of the paraboloid from the spherical surface.
In the above polar coordinates, the radial deviation $d$  of the paraboloid from the spherical surface can be expressed as,

\begin{equation}
    d=|\rho_s-\rho_p|,
	\label{eq:eq4}
\end{equation}
 When the vertex of the paraboloid touches the spherical surface, the paraboloid has minimum  $d$ at its vertex. The paraboloid is in the specified position in polar coordinates, the polar angle corresponding to the vertex of the paraboloid is $\frac{3}{2}\pi$, and formula ~(\ref{eq:eq5}) can be obtained as follows,

\begin{equation}
    c=R^2\cos^2(\frac{3}{2}\pi)-2pR\sin(\frac{3}{2}\pi).
	\label{eq:eq5}
\end{equation}

Here the paraboloid conforms to both formula ~(\ref{eq:eq3}) and formula ~(\ref{eq:eq5}),  $R$ and $D$ are known, the paraboloid can be obtained and the radial deviation of the paraboloid (within its polar angle range from the spherical surface)can be calculated.

(2)	The paraboloid with the minimum $d_{max}$  is selected \citep{article}, where $d_{max}$ is the maximum radial deviation of the paraboloid within its polar angle range from the spherical surface.

As described in condition (1), the radial deviation of the paraboloid from the spherical surface is denoted by $d$ in meters. According to the relationship between the paraboloid and the spherical surface (as shown by formula ~(\ref{eq:eq1}) and formula ~(\ref{eq:eq2})), the radial deviation $d$  is different within the polar angle range of the paraboloid, the maximum deviation being defined as $d_{max}$. We then select the paraboloid with the minimum $d_{max}$.

Obviously, a smaller $d_{max}$ is desirable for adjusting the active reflector of the telescope, so condition (2) can be taken as the recommended approach for selection of the paraboloid of the active reflector in general.



\section{Analysis of the paraboloid of FAST}

The huge FAST reflector of 500\,m diameter is divided into 4300 triangular panels. Each panel, with edges roughly 11 m long, has a spherical surface with a 315\,m curvature radius and a shape error of about 2\,mm RMS. The parabolic sector is geometrically fitted by about 1400 triangular panels.

If one wants a significant bigger (or smaller) aperture than 300\,m (or to go to higher frequencies), one has to increase (or decrease) the travel lengths of hydraulic actuators, while reducing (or enlarging) the size of triangular panels to keep the fitting accuracy of the reflector surface because of the inherent geometrical difference between spherical and parabolic surfaces, which will be the topic of future discussion on applying the active reflector technology to an expanded array of FAST like telescopes.

\begin{figure}
	\includegraphics[width=\columnwidth]{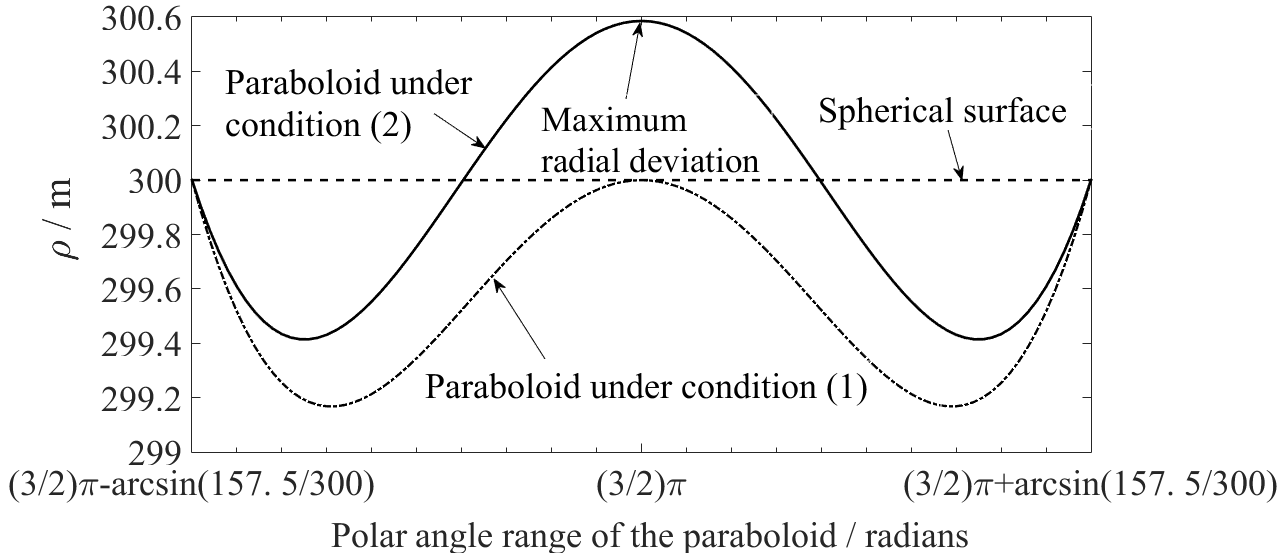}
    \caption{Radial deviations of the paraboloids from the spherical surface. The abscissa represents the polar angle range corresponding to a paraboloid of 315\,m aperture in radians. The ordinate shows the radial length from the pole to the spherical surface in meters, and it also shows the radial length from the pole to the paraboloids under the two calculation conditions, in meters. The dashed line indicates the radial length from the pole to the spherical surface. The dash-dotted line indicates the radial length from the pole to the paraboloid under condition (1) and the solid line indicates the radial length from the pole to the paraboloid under condition (2). As mentioned in section 2,condition(1) requires the radial distance of the paraboloid from the spherical surface at the parabolic vertex to have a minimum value; condition (2) requires that the maximum radial distance of the paraboloid from the spherical surface is a minimum. }
    \label{fig:fig4_figure}
\end{figure}

The standard aperture of the FAST paraboloid is R=300\,m. If the aperture can be made larger than 300\,m, the sensitivity of FAST will improve. A paraboloid of 315\,m aperture can be analyzed as follows:

If the paraboloid is in the specified position in polar coordinates as mentioned in section 2, the corresponding polar angle range would need to be $(\frac{3}{2}\pi-\arcsin(\frac{157.5}{300}),\frac{3}{2}\pi+\arcsin(\frac{157.5}{300}) )$. The paraboloid can be obtained under two conditions described in section 2. The radial deviations of the paraboloids from the spherical surface are shown in Figure ~\ref{fig:fig4_figure}.The radial deviations of the paraboloid from the spherical surface can be obtained for each. As shown in Figure ~\ref{fig:fig4_figure}, the maximum radial deviation of the paraboloid from the spherical surface corresponding to condition (2) is about 0.59\,m.

\begin{figure}
	\includegraphics[width=\columnwidth]{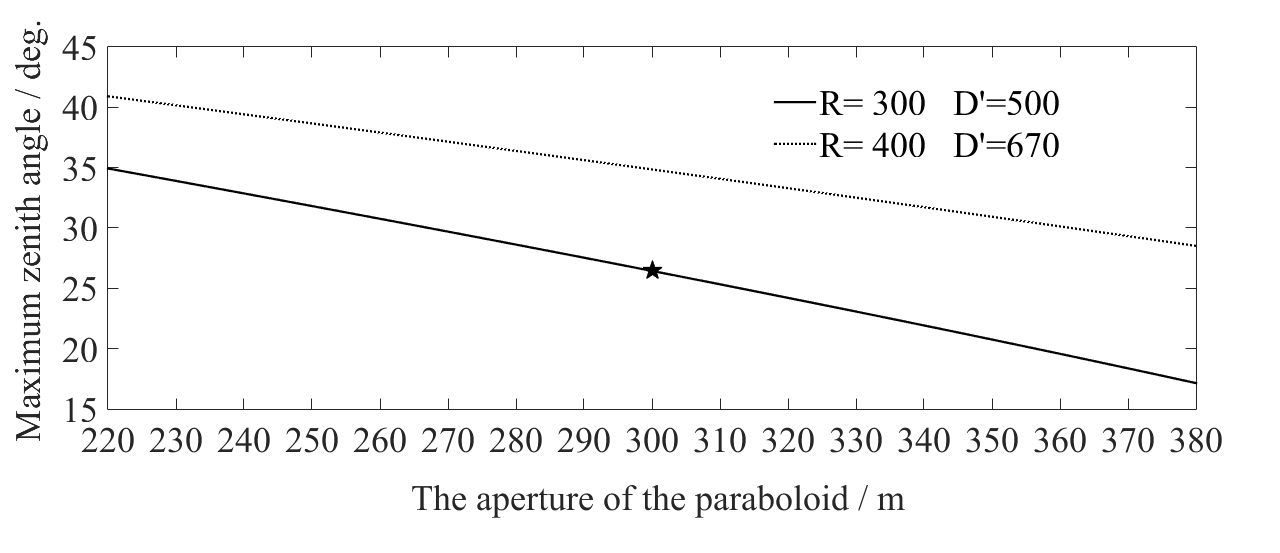}
    \caption{The maximum zenith angle corresponding to the apertures of the paraboloid. The abscissa represents the aperture of the paraboloid in meters. The ordinate shows the maximum zenith angle corresponding to the apertures of the paraboloid in degrees.
    The solid line is for the spherical surface of FAST, whose radius is 300\,m and opening aperture is 500\,m. The dotted line is for the spherical surface with a radius of 400\,m and an opening aperture of 670\,m. A star indicates the maximum zenith angle of the 300\,m aperture of FAST, which is about 26.4 deg.}
    \label{fig:fig5_figure}
\end{figure}

For FAST, the opening aperture $D'$  of the main spherical surface is 500\,m. The paraboloid is kept within the main spherical surface and does not extend beyond the edge of the primary spherical surface. We need to consider also that the actuator maximum travel is about $\pm$0.6\,m  (\citealt*{2017Initial}; \citealt{2017Primary}),which is the design requirement and is also achieved in practice. For the main spherical surface with a radius of 300\,m and an opening aperture of 500\,m, the maximum zenith angle corresponding to the paraboloid apertures is shown in Figure ~\ref{fig:fig5_figure}(solid line).If the aperture of the paraboloid is decreased to 220\,m, which is similar to the illuminated aperture of the Arecibo 305\,m radio telescope \citep{1995ASPC...75...90G}, a larger zenith angle can be obtained.It is the Gregorian system of Arecibo that limits the illumination of the primary to 220 m. Non-imaging line feeds can illuminate the whole reflector. When the paraboloid corresponds to the minimum $d_{max}$ , for the 300\,m aperture of FAST, the calculated maximum radial deviation is about 0.47\,m, which allows for a comfortable margin compared to the actuator maximum travel of 0.6\,m. Should the aperture of the paraboloid be raised to 315\,m, the calculated maximum radial deviation is close to the actuator maximum travel, the increased effective collecting area can improve the sensitivity of the telescope (at the cost of decreasing the maximum zenith angle, of course).

\section{Analysis of the paraboloid and spherical surface of different size}
Radio telescopes of large effective collecting area are essential to the development of radio astronomy. Apertures larger than those of existing radio telescopes can be analyzed, but at the same time the aperture of the paraboloid is also limited by factors such as the site of the telescope and engineering costs. Different apertures for the paraboloid can meet the scientific goals of large sky coverage or high sensitivity. The application of active reflector technology to FAST-like telescopes could be optimized to obtain paraboloids of larger aperture for higher sensitivity, or to achieve greater zenith angles for increased sky coverage.

We take 300\,m$\pm$80\,m  for the range of the paraboloid aperture. At the lower end, a paraboloid of 220\,m aperture has sensitivity similar to the Arecibo 305\,m radio telescope. Reference data for the active reflector of FAST-like telescopes can be provided by analyzing the main spherical surface using different radii. Spherical surfaces with radii of 300\,m and 400\,m are considered. The analysis of paraboloids of the given apertures for a spherical radius of 300\,m can be compared with the analysis for a spherical radius of 400\,m. The achievable zenith angles for paraboloids of different apertures are also analyzed.

\begin{figure}
	\includegraphics[width=\columnwidth]{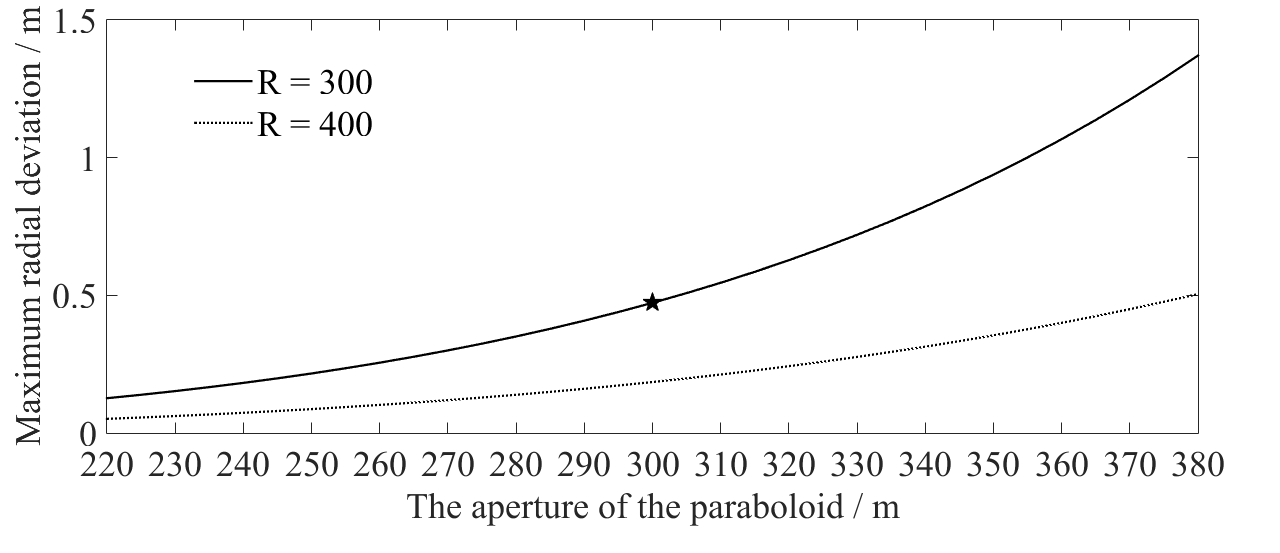}
    \caption{The calculated maximum radial deviation of the paraboloid from the spherical surface. The abscissa represents the aperture of the paraboloid in meters. The ordinate shows the calculated maximum radial deviation of the paraboloid from the spherical surface in meters. The solid line indicates the calculated maximum radial deviation when the radius of the main spherical surface is 300\,m. The dotted line is for the calculated maximum radial deviation when the radius of the main spherical surface is 400\,m. A star indicates that the maximum radial deviation of the 300\,m aperture of FAST is about 0.47\,m.}
    \label{fig:fig6_figure}
\end{figure}

Figure ~\ref{fig:fig6_figure} shows the maximum radial deviation of the paraboloids of different apertures from the spherical surface for radii of 300\,m and 400\,m. For each given aperture, the calculated maximum radial deviation is that for the paraboloid that has the smallest $d_{max}$. Larger diameter telescopes appear to require less actuator motion to achieve the parabola, which is indicated in Figure 6.

If larger sky coverage is required, the aperture of the paraboloid can be decreased and the opening aperture of the main spherical surface could be increased. For the spherical radius of 400\,m, we analyzed the zenith angle of a spherical reflector, whose $\frac{R}{D'}$ ratio (about 0.597) between the spherical radius and the spherical opening aperture is similar to that (0.6) of FAST. We used 670\,m as the opening aperture of the spherical surface for analysis of the zenith angle. The paraboloid is kept within the main spherical surface and does not extend beyond the edge of that surface; the maximum zenith angle corresponding to the paraboloid apertures is shown in Figure ~\ref{fig:fig5_figure}(dotted line).

\section{Discussion and summary}

For FAST, by analyzing radial deviations of the paraboloid from the primary spherical reflector, possible different apertures for the paraboloid are obtained to explore potential improvement in the performance of the telescope. If the aperture of the FAST paraboloid can be increased or decreased from its current size, different observing modes of the telescope can be enabled. For example, a tracking mode could operate with a decreased aperture, while a drift-scan mode could be implemented with an increased aperture on the premise of meeting diverse scientific goals.

A large number of karst depressions, more than 400, have been found in the south of Guizhou \citep{1998IAUS..179...93P}. These provide feasible sites for the proposed FAST-like telescopes and eventually the telescope array. The array will further greatly improve both sensitivity and angular resolution for radio astronomy. The analysis of the application of active reflector technology to FAST-like telescopes in this paper will provide technical support for the telescope array.

a. During the analysis of FAST paraboloids of different apertures, the apertures increased to 315\,m. When the paraboloid corresponds to the minimum $d_{max}$  , the calculated maximum radial deviation is about 0.59\,m. This is theoretically feasible for the present actuator travel range, and it is close to the current actuator maximum travel (about $\pm$0.6\,m ). For the spherical radius of 400\,m, during the analysis of paraboloids of the given apertures of 300\,m$\pm$80\,m, corresponding to a paraboloid aperture of 380\,m, similarly, the calculated maximum radial deviation is about 0.51\,m. The corresponding maximum radial deviation increases with increasing aperture of the paraboloid.

b. The achievable zenith angle is analyzed assuming the paraboloid is kept within the main spherical surface and does not extend over its edge. For a spherical surface of radius 400\,m and opening aperture of 670\,m, the maximum zenith angle is about 41 degrees if the aperture of the paraboloid is 220\,m. The maximum zenith angle increases with decreasing paraboloid aperture.

c. If the aperture of the paraboloid changes, the feed cabin suspension and the positioning system may need to be reanalyzed. The feed illumination will be changed accordingly to match the aperture of the paraboloid, so there will be a match between feed and reflector.

FAST is the world\arcmin s largest single dish radio telescope, the application of active reflector technology to FAST is one of its outstanding features. If the aperture of the paraboloid of FAST increases from 300\,m to 315\,m, the sensitivity of the telescope will be improved by approximately 10\%. Assuming the paraboloid is within the main spherical surface and does not extend beyond its edge, if its aperture decreases to 220\,m (which is similar to the illuminated aperture of the Arecibo 305\,m radio telescope), the maximum zenith angle will increase to approximately 35 degrees compared to the 26.4 degrees of current FAST.

New feed horns are to be designed and manufactured in the future for upgrades, to match a larger or smaller diameter surface with the appropriate radiation pattern. The current FAST receiver platform, where any feed horn can be quickly switched in and out of the telescope focus in a minute, can accommodate any new feeds required. One special feature of FAST is that outside of the deformed 300\,m aperture  parabolic reflector,there are large areas of the spherical surface that efficiently block thermal radiation from the ground. A single feed with less edge taper at 300\,m could be used at FAST, and could potentially accommodate a modest range of apertures that deviate by <$\sim 5\%$ from 300\,m. Currently, we are developing a Cryo-PAF(Phased Array Feed) for the FAST telescope. The PAF would be able to accommodate a range of f/D ratios, corresponding to different apertures of the deformed area.

The control algorithm will be further developed to transfer among different horns as required, which is needed for the dish deformation. Different apertures impose different deformation requirements, as do the stress and fatigue of the cable-net which are closely related to the structural  integrity of the telescope.

Considering the frequency coverage of FAST, most of the scientific targets are point like sources with respect to the FAST beam-size of about 3 arc minutes at L band, which should not be affected largely by the beam-size varying. The narrower beam resulting from a larger aperture would certainly have some impact on survey observations. Some modifications of the pointing arrangement would be needed.

The increasing and decreasing of the aperture of the paraboloid in the active reflector would enable the telescope to meet various scientific goals, e.g., deeper drift surveys and/or more sky coverage. This study of spherical surfaces of different size could provide reference data for active reflectors of FAST-like telescopes. We look forward to the application of active reflector technology to FAST-like telescopes, to create more research possibilities in the future.

\section*{Acknowledgements}

This work is supported by the National Key $R \& D$ Program of China under grant number 2018YFA0404703, and the Open Project Program of the CAS Key Laboratory of FAST, NAOC, Chinese Academy of Sciences.The authors thank all the staff of JLRAT, National Astronomical Observatories, Chinese Academy of Sciences.RGS thanks NAOC and the CAS for hospitality and (financial) support during countless visits to the PRC.

\section*{Data Availability}

The data underlying this article will be shared on reasonable request to the corresponding author.



\bibliographystyle{mnras}
\bibliography{bibtex} 





\bsp	
\label{lastpage}
\end{document}